\documentclass[%
 reprint,
superscriptaddress,
 amsmath,amssymb,
 aps,
]{revtex4-2}

\usepackage{graphicx}
\usepackage{dcolumn}
\usepackage{bm}

\usepackage{xcolor}

\begin{document}

\preprint{APS/123-QED}


\title{Multimessenger Prospects for Low-Luminosity Gamma-Ray Bursts: Joint Neutrino and X-Ray Observations}

\author{Wenkang Lian}
\affiliation{School of Physics and Astronomy, Beijing Normal University, Beijing 100875, People’s Republic of China}
\affiliation{
Institute of High Energy Physics, Chinese Academy of Sciences}

\author{He Gao}
\email{gaohe@bnu.edu.cn}
\affiliation{School of Physics and Astronomy, Beijing Normal University, Beijing 100875, People’s Republic of China}
\affiliation{Institute for Frontier in Astronomy and Astrophysics, Beijing Normal University, Beijing 102206, People’s Republic of China}

\author{Shunke Ai}
\affiliation{Niels Bohr International Academy and DARK, Niels Bohr Institute, University of Copenhagen, Blegdamsvej 17, 2100 Copenhagen, Denmark}

\author{B. Theodore Zhang}
\affiliation{Key Laboratory of Particle Astrophysics and Experimental Physics Division and Computing Center, Institute of High Energy Physics, Chinese Academy of Sciences, 100049 Beijing, China}
\affiliation{TIANFU Cosmic Ray Research Center, 610213,Chengdu, Sichuan , China}





\date{\today}

\begin{abstract}
Low--luminosity gamma-ray bursts (LLGRBs) are promising candidates for high-energy neutrinos, yet no coincident neutrino events have been detected so far. 
Recent advances in X-ray time-domain astronomy, together with the development of next-generation neutrino telescopes, open new opportunities for joint X-ray and neutrino observations of these transients.
We calculate the jet dynamical evolution and the associated neutrino production for both non-magnetized and magnetized outflows. 
For individual events, joint X-ray and neutrino detection is generally limited to nearby LLGRBs or sources with high luminosities.
Thus, we consider a next-generation neutrino telescope with an effective area enhanced by a factor of $\sim30$ relative to IceCube. 
In the non-magnetized scenario, joint detection of individual events is enabled for sources with typical isotropic luminosities of $L_{\mathrm{iso}}\sim10^{47}\,\mathrm{erg\,s^{-1}}$ out to luminosity distances of $D_L\sim1.6\times10^{2}\,\mathrm{Mpc}$, corresponding to an expected detection rate of order $1$ per year.
In contrast, for the magnetized scenario at the same luminosity, the accessible distance is significantly reduced, with joint observations confined to sources within $D_L\sim6.5\times10^{1}\,\mathrm{Mpc}$ and an expected detection rate of order $0.5$ per year.
For stacked samples of $\sim100$ magnetized LLGRBs, stacking substantially enlarges the accessible distance range, enabling joint observations for sources with representative luminosities of $L_{\mathrm{iso}}\sim1\times10^{47}\,\mathrm{erg\,s^{-1}}$ out to $D_L\lesssim7.0\times10^{2}\,\mathrm{Mpc}$ and corresponding to an expected detection rate of order $2$ per year.
These results demonstrate that joint X-ray and next-generation neutrino observations enable a practical multimessenger probe of LLGRBs.


\end{abstract}

\maketitle

\section{\label{sec:level1}INTRODUCTION}

Gamma-ray bursts (GRBs) are thought to arise from ultra-relativistic outflows produced either in the collapse of massive stars~\cite{1993ApJ...405..273W, 1998Natur.395..670G, 2003ApJ...586..356Z, 2003Natur.423..847H, 2003ApJ...591L..17S} or in the mergers of compact binary systems~\cite{1986ApJ...308L..43P, 1989Natur.340..126E, 1992ApJ...395L..83N, 2017PhRvL.119p1101A}.
These outflows provide natural sites for high-energy particle acceleration.
Thus, GRBs have long been discussed as possible accelerators of ultrahigh-energy cosmic rays (UHECRs)~\cite{1995ApJ...453..883V, 1995PhRvL..75..386W} and as potential sources of very high-energy neutrinos~\cite{1997PhRvL..78.2292W, 
refId0, 2018A&A...611A.101B, 2022ApJ...941L..10M}.

Despite more than a decade of IceCube observations, no significant neutrino signal has been found in temporal and directional coincidence with any GRB~\cite{2011PhRvL.106n1101A,  2017ApJ...843..112A, 2022ApJ...939..116A, 2023A&A...672A.102L}.
Even for GRB~221009A, dubbed the “brightest of all time”~\cite{2023ApJ...946L..31B}, no associated neutrinos have been detected~\cite{2023ApJ...944..115A, 2023ApJ...946L..26A, 2024JCAP...08..006A, 2025ApJ...987...79L}.
Current limits indicate that high-luminosity GRBs (HLGRBs) contribute only about $\sim 1\%$ to the diffuse astrophysical neutrino flux~\cite{2015ApJ...805L...5A}.
This motivates a closer look at other GRB populations and environments where neutrino production may be more efficient.

A particularly interesting class is that of low-luminosity GRBs (LLGRBs).
These events have relatively low isotropic-equivalent luminosities, $L_{\mathrm{iso}} \lesssim 10^{49}\,\mathrm{erg\,s^{-1}}$, and long durations of several thousand seconds~\cite{2007ApJ...662.1111L, PhysRevD.97.083010}.
Because their event rate is higher than that of HLGRBs, LLGRBs have been proposed as promising contributors to the diffuse neutrino background~\cite{2006ApJ...651L...5M, 2016PhRvD..93h3003S}.
GRB~060218 is a well-studied example of this class.
Its association with SN~2006aj firmly points to the core collapse of a massive star~\cite{2006Natur.442.1008C}.
Ref.~\cite{2015ApJ...807..172N} interpreted GRB~060218 as a shock-breakout event and proposed a unified picture that links HLGRBs and LLGRBs.
In this picture both classes share similar progenitors, but LLGRBs are embedded in dense extended material that slows or halts the jet.

The recently observed event EP240414A also suggests that dense extended material may surround the progenitor before core collapse~\cite{2025NatAs...9.1073S, 2025ApJ...986L...4H}.
Several works have pointed out that in such environments the jet can be fully choked and produce only neutrinos without prompt electromagnetic counterparts, often called orphan neutrinos~\cite{2016PhRvD..93h3003S, Denton_2018, 2018ApJ...856..119H, 2021ApJ...911L..19Z}.

Dense extended material naturally enhances hadronic interactions.
As the relativistic jet propagates through this material, protons accelerated at the forward and reverse shocks can undergo $pp$ and $p\gamma$ interactions, which create pions and kaons that then decay into high-energy neutrinos~\cite{1997PhRvL..78.2292W,Denton_2018,2021ApJ...911L..19Z}.
The overall efficiency of neutrino production depends on the jet luminosity, Lorentz factor, and the density profile of the surrounding material~\cite{2013PhRvL.111l1102M,2016PhRvD..93h3003S,2018ApJ...856..119H}, and may also be affected by the jet magnetization.

The rapid advances in X-ray time-domain astronomy and in neutrino telescope technology now makes it possible to test these ideas.
The \textit{Einstein Probe} (EP)~\cite{2025SCPMA..6839501Y}, with its Wide-field X-ray Telescope(WXT), is expected to detect nearby LLGRBs whose emission peaks in the X-ray band.
At the same time, several next-generation neutrino telescopes, such as IceCube-Gen2~\cite{2021JPhG...48f0501A}, KM3NeT~\cite{2016JPhG...43h4001A}, TRIDENT~\cite{2023NatAs...7.1497Y} and the Hunt experiment~\cite{2024icrc.confE1080H}, are being built or proposed.
These facilities will have much better sensitivity to astrophysical neutrinos and will enable joint searches for LLGRBs and their neutrino counterparts.

In this work, we investigate the interaction between a relativistic jet and dense extended material around the progenitor.
We follow the formation and evolution of the forward and reverse shocks and compute the corresponding internal energy densities.
Using these results, we calculate the efficiency of neutrino production, including all relevant cooling processes.
Based on the dynamical evolution, we classify the outcomes into three regimes: choked-jet events, HLGRBs, and LLGRBs.
We use the observed LLGRB-to-HLGRB rate ratio to infer the intrinsic rate of choked-jet event.
This allows us to estimate the diffuse neutrino flux from both LLGRBs and choked-jet events.

Most importantly, we quantify the parameter space in the $(L_{\rm iso}, D_L)$ plane, with $L_{\rm iso}$ being the isotropic-equivalent luminosity and $D_L$ the luminosity distance, to assess the detectability of individual sources with neutrino telescopes and X-ray detectors, as well as the effectiveness of stacking $\sim 100$ sources.
We perform this analysis for both non-magnetized and magnetized jets and discuss the capabilities of current and future neutrino telescopes.

\section{Theoretical Model}
\label{sec:theory}

Here we assume that the progenitors of both low-luminosity and high-luminosity gamma-ray bursts are Wolf–Rayet stars, while the key difference lies in the circumstellar environment. Low-luminosity events are embedded in dense extended material that can slow or choke the jet~\cite{2015ApJ...807..172N}.

\subsection{Jet Propagation Dynamics}
\label{sec:jet_dynamics}

\begin{figure}[htbp]
    \centering
    \includegraphics[width=0.45\textwidth]{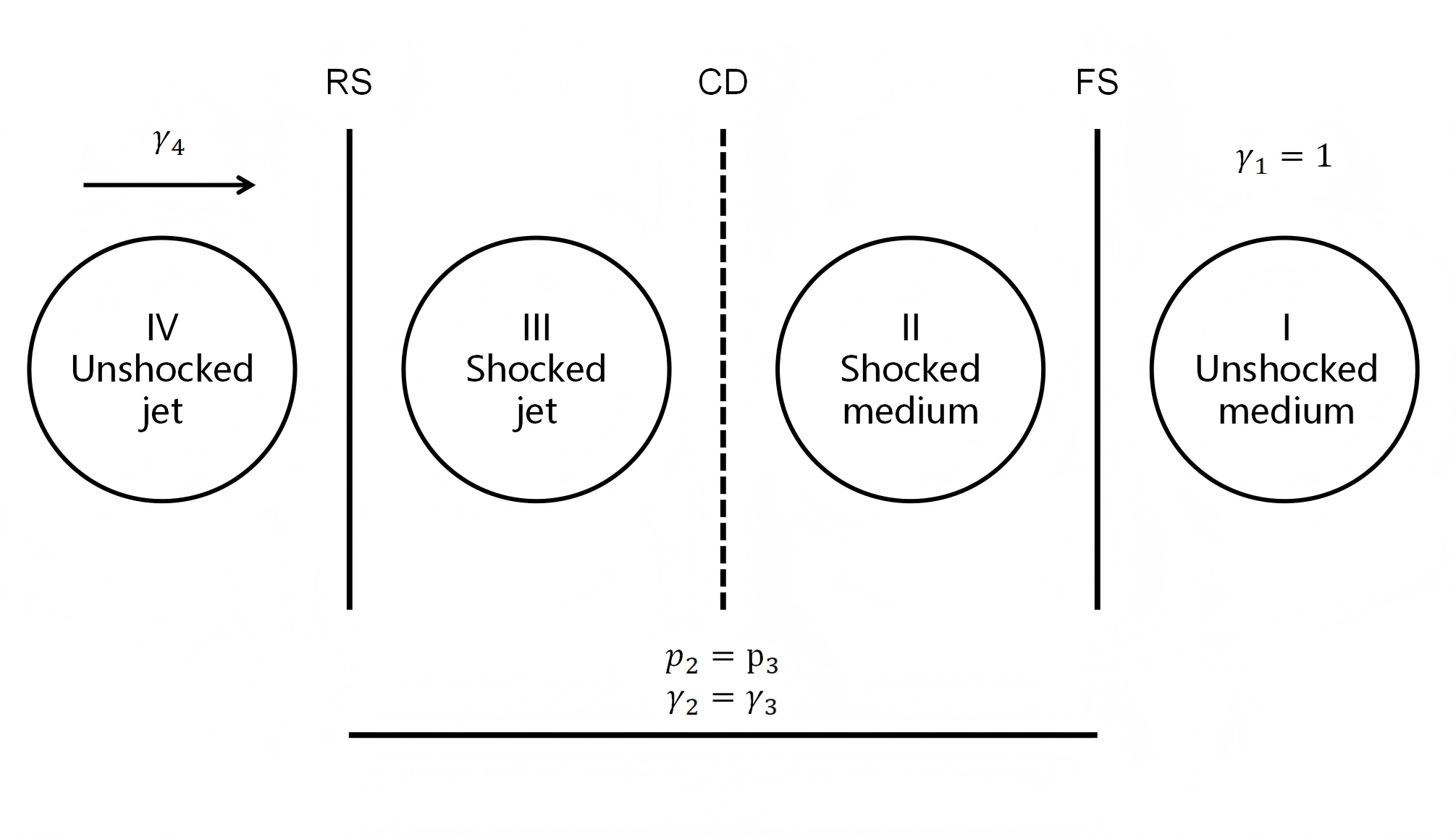}
    \caption{
    Structure of the FS--RS system. Regions I--IV are separated by the FS, CD, and the RS.}
    \label{fig:fs_rs_structure}
\end{figure}

After emerging from the stellar surface, the jet becomes uncollimated and accelerates to relativistic speeds~\cite{Bromberg_2011}.
Its interaction with the surrounding material gives rise to a forward shock (FS) propagating into the extended-material and a reverse shock (RS) propagating into the jet~\cite{2003MNRAS.345..575M,Yu_2020}.
The region bounded by the FS and RS defines the jet head.
\begin{figure}[t]
    \centering
    \includegraphics[width=0.8\linewidth]{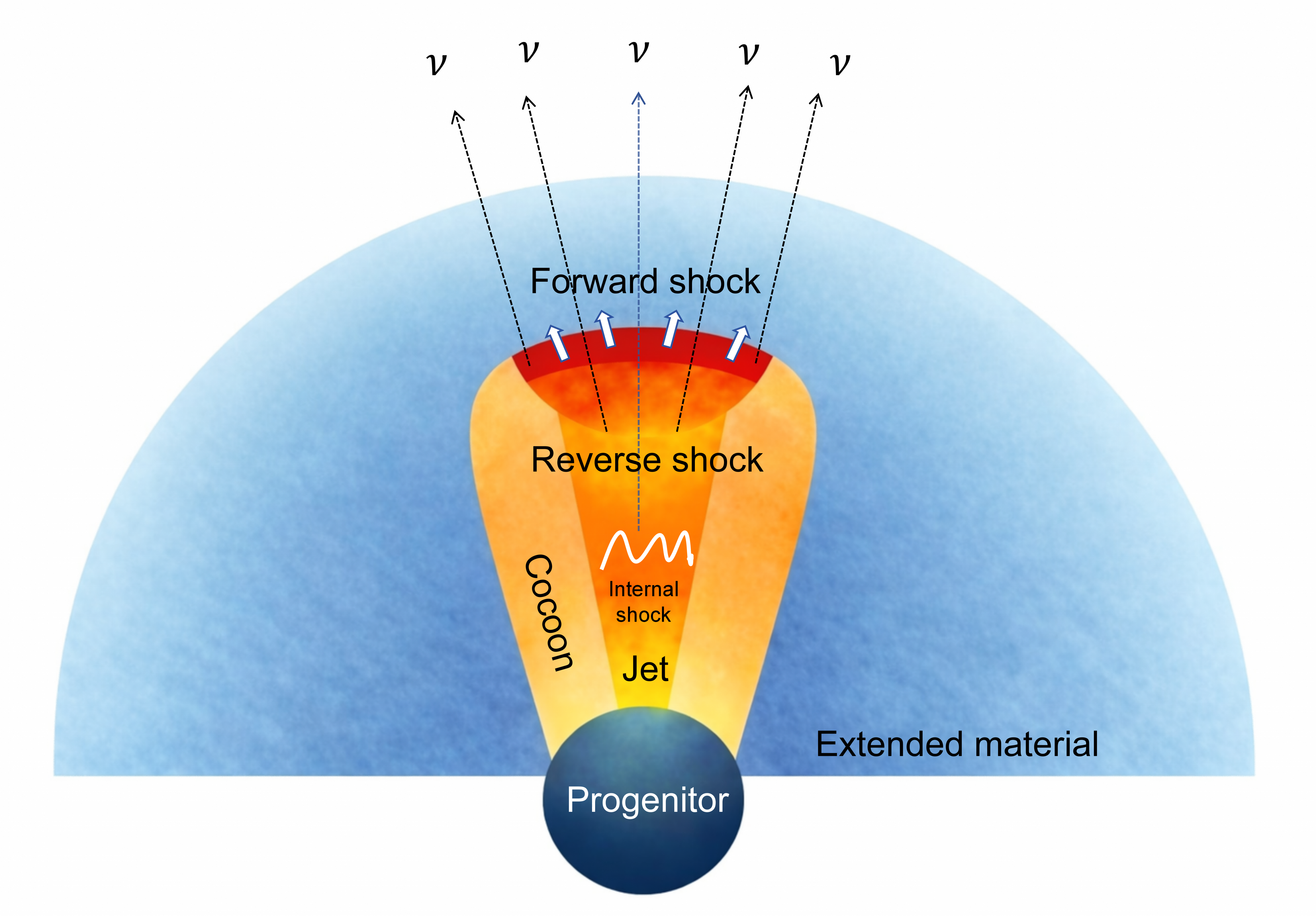}
    \caption{
    Schematic illustration of the jet–extended-material interaction.
    }
    \label{fig:jet_dynamics}
\end{figure}
A schematic illustration of this jet–extended-material interaction is shown in Fig.~\ref{fig:jet_dynamics}. A more detailed view of the jet-head structure is shown in Fig.~\ref{fig:fs_rs_structure}.

As illustrated in Fig.~\ref{fig:fs_rs_structure}, Region~I corresponds to the unshocked ambient medium, Region~II to the shocked ambient medium, Region~III to the shocked jet material, and Region~IV to the unshocked jet.
The FS separates Regions~I and II, while the RS separates Regions~III and IV.
The contact discontinuity (CD) lies between Regions~II and III.
Throughout this work, $Q_{\chi\eta}$ denotes the value of the quantity $Q$ in region $\chi$ measured in the rest frame of region $\eta$, and $Q_{\chi}$ denotes the value of the quantity $Q$ in region $\chi$ measured in its own rest frame. 
For velocity-dependent quantities, if only a single subscript is used, the quantity is defined in the lab rest frame.

We assume cold upstream flows and adopt $\gamma_2 = \gamma_3$ and $p_2 = p_3$ in the shocked regions.
The jet is characterized by its intrinsic luminosity $L_0$, bulk Lorentz factor $\Gamma_j$, and magnetization $\sigma_4$.
The comoving proton number density in the unshocked jet (Region~IV) is given by
$n_4 = {L_0}/{(4\pi \Gamma_j^{2} r_{\text{rs}}^{2} m_p c^{3} (1+\sigma_4))}. $
Here, $r$ denotes the radial coordinate, while $r_{\text{fs}}$ and $r_{\text{rs}}$ correspond to the radii of the forward and reverse shocks, respectively.

Following Ref.~\cite{2005ApJ...628..315Z}, assuming $\gamma_1 = 1$, the number densities and internal energy densities in the shocked regions satisfy
\begin{equation}
\begin{aligned}
\frac{e_2}{n_2 m_p c^2} &= (\gamma_2 - 1), \\
\frac{n_2}{n_1} &= 4\gamma_2 + 3, \\
\frac{e_3}{n_3 m_p c^2} &= (\gamma_{34} - 1) f_a, \\
\frac{n_3}{n_4} &= (4\gamma_{34} + 3) f_b ,
\end{aligned}
\end{equation}
where $\gamma_{34}$ is the relative Lorentz factor between Regions~III and IV, and $f_a$ and $f_b$ are correction factors defined in Ref.~\cite{2005ApJ...628..315Z}.

Using the equation of state $p = (\hat{\Gamma}-1)e$ with $\hat{\Gamma}=4/3$ appropriate for a relativistic downstream, the pressures in the shocked regions can be expressed in terms of $e_2$ and $e_3$.
Imposing pressure balance across the contact discontinuity, $p_2 = p_3$, allows the Lorentz factors of the jet head to be solved once the external density profile $\rho(r_{\text{fs}})$ is specified.

In the magnetized case, Region~III receives an additional pressure contribution from the magnetic field carried by the jet.
The magnetic field strength in Region~III satisfies
$B_3^{2} = 4\pi n_3 \mu_3 \sigma_3$,
where $\mu_3$ is the specific enthalpy,
$\mu = m_p c^{2} + \frac{\hat{\Gamma}}{\hat{\Gamma}-1}\frac{p}{n}$.
This magnetic contribution modifies the pressure balance condition and affects the resulting shock dynamics.

Once the bulk Lorentz factor of the jet head, $\Gamma_h$, is determined, the velocity of the reverse shock follows from Eq.~(A16) of Ref.~\cite{2005ApJ...628..315Z}.
The jet becomes choked when the reverse shock has fully crossed the jet.

\begin{figure}[htbp]
    \centering
    \includegraphics[width=0.45\textwidth]{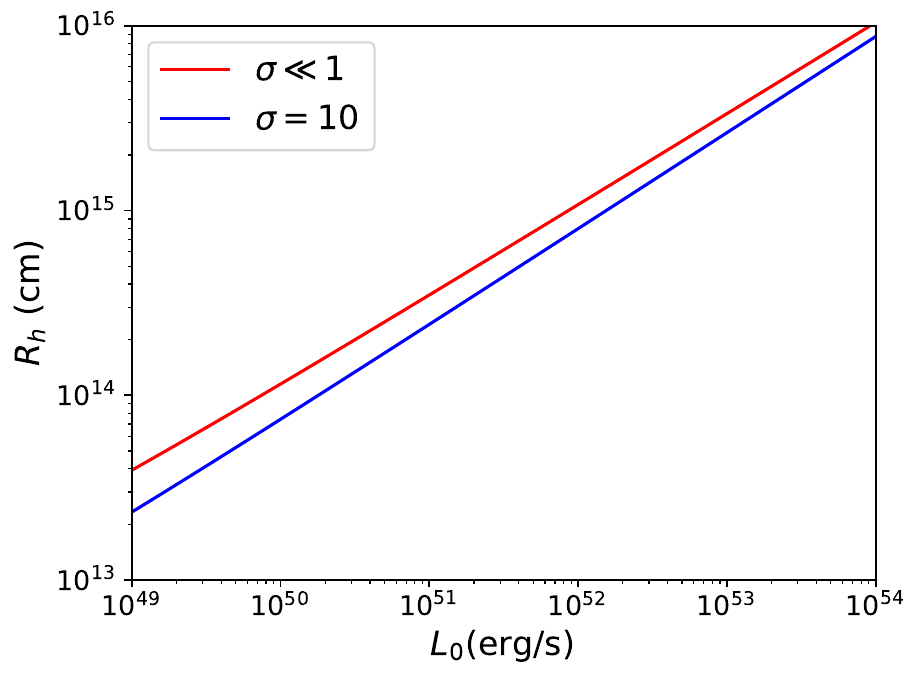}
    \caption{
    Jet head radius $R_{\mathrm{h}}$ at which the jet becomes choked as a function of the intrinsic luminosity $L_0$.
    The results are shown for different magnetization parameters.
    }
    \label{fig:l_rchock}
\end{figure}

To illustrate the jet dynamics, we adopt a bulk Lorentz factor determined by the empirical $\Gamma_{\mathrm{j}}$–$L_0$ correlation~\cite{2010ApJ...725.2209L, 2012ApJ...751...49L, PhysRevLett.110.121101},
\begin{equation}
    \Gamma_{\mathrm{j}} \simeq 250\,L_{0,52}^{0.30},
\end{equation}
where $L_{0,52} = L_0 / 10^{52}\,\mathrm{erg\,s^{-1}}$. We fix the engine duration to $30\,\mathrm{s}$ and adopt an external density profile motivated by GRB~060218.
Following Ref.~\cite{2007ApJ...667..351W}, the shock breakout of GRB~060218 occurs at $r_{\rm sb}=7.8\times10^{12}\,\mathrm{cm}$, where the density is $\rho(r_{\rm sb})\approx10^{-12}\,\mathrm{g\,cm^{-3}}$.
Assuming a wind-like profile, we write $\rho(r)=A r^{-2}$ with $A\simeq6\times10^{13}\,\mathrm{g\,cm^{-1}}$.
Using this density profile, the jet head radius at which the jet is finally choked, $R_{\mathrm{h}}$, is shown for different intrinsic luminosities $L_0$ in Fig.~\ref{fig:l_rchock}.
We can see that choking occurs more easily in magnetized jets.

For $L_0 = 10^{49}\,\mathrm{erg\,s^{-1}}$, the jet is choked at $R_{\mathrm{h}} \approx 3.8\times10^{13}\,\mathrm{cm}$ in the non-magnetized case ($\sigma \ll 1$), whereas in the magnetized case ($\sigma=10$) the choking radius is reduced to $R_{\mathrm{h}} \approx 2.3\times10^{13}\,\mathrm{cm}$.
The isotropic shock-breakout luminosity is estimated as
$L_{\mathrm{iso}} \simeq 4\pi r_{\mathrm{sb}}^{2}\beta_{\mathrm{sb}} c\,e_2\,\xi_e \simeq 10^{46}\,\mathrm{erg\,s^{-1}}$,
where $\beta_{\mathrm{sb}}$ is the breakout velocity in units of $c$ and $\xi_e \sim 0.1$ is the electron energy fraction, consistent with GRB~060218~\cite{2006Natur.442.1008C}.

\subsection{Neutrino Emission}
\label{sec:neutrino_emission}

Once the jet dynamics are known, the neutrino production rate at each radius can be calculated.

We first consider neutrino production in the FS--RS system.  
Protons can be accelerated at the RS only when the radiation constraint is satisfied~\cite{2013PhRvL.111l1102M},
\begin{equation}
\tau_{\mathrm{T}} = n_{4}\sigma_{\mathrm{T}} r_{\text{rs}} / \Gamma_{j}
\lesssim \min \left[ 1,\; 0.1\, C^{-1}\gamma_{34} \right],
\label{eq:tauT_RS}
\end{equation}
where $\sigma_{\mathrm{T}}$ is the Thomson cross section, and $C = 1 + 2 \ln(\gamma_{34}^{2})$.
Proton acceleration in the forward shock is assumed to occur only after shock breakout, which for a relativistic shock corresponds to the stage when the Thomson optical depth ahead of the shock drops below unity.

The proton acceleration timescale is $t_{p,\mathrm{acc}} = \eta \epsilon_{p} / (e B c)$,
where $\epsilon_p$ is the proton energy, $B$ is the magnetic field strength, and $\eta = 1$ is adopted, corresponding to the Bohm diffusion case. 
For a non-magnetized downstream ($\sigma_3 \ll 1$), we take $B_3 = (8\pi \xi_B e_3)^{1/2}$ with $\xi_B = 0.1$.  
For a magnetized jet, $B_3$ includes an additional magnetic contribution originating from the jet itself, with $B_3^{2} = 4\pi n_3 \mu_3 \sigma_3$.

In the jet head we consider only thermal photons, since the number densities of other components such as keV--MeV prompt photons are much lower~\cite{2016PhRvD..93h3003S,2021ApJ...911L..19Z}.  
The temperature of thermal photons in Regions II and III is estimated as 
\begin{equation}
k_{\mathrm{B}}T_{2(3)}
= \left(\frac{15\,\hbar^{3}c^{3}\xi_{e}\,e_{2(3)}}{\pi^{2}}\right)^{1/4},
\end{equation}
where $\hbar$ is the reduced Planck constant.

A high-energy proton cools through synchrotron radiation, inverse-Compton (IC) scattering, hadronic interactions, and adiabatic expansion.  
The synchrotron cooling timescale is~\cite{1979rpa..book.....R}
\begin{equation}
t_{p,\mathrm{syn}}
= \frac{6\pi m_p^{4} c^{3}}{\sigma_{\mathrm{T}} m_e^{2} B^{2} \epsilon_p},
\label{eq:t_syn}
\end{equation}
where $m_e$ is the electron mass.

The IC cooling timescale is~\cite{1979rpa..book.....R}
\begin{equation}
t_{p,\mathrm{IC}} =
\begin{cases}
\displaystyle \frac{3 m_p^{4} c^{3}}{4\sigma_{\mathrm{T}} m_e^{2} n_{\gamma} \langle \epsilon_{\gamma} \rangle \epsilon_p},
& \langle \epsilon_{\gamma} \rangle \epsilon_p < m_p^{2} c^{4}, \\[6pt]
\displaystyle \frac{3 \langle \epsilon_{\gamma} \rangle \epsilon_p}{4\sigma_{\mathrm{T}} m_e^{2} c^{5} n_{\gamma}},
& \langle \epsilon_{\gamma} \rangle \epsilon_p > m_p^{2} c^{4},
\end{cases}
\label{eq:t_IC}
\end{equation}
where $n_{\gamma}$ and $\langle \epsilon_\gamma \rangle = 2.7\,k_B T_{2(3)}$ are the number density and the average energy of target photons.

Protons also lose energy through hadronic channels, including  
photomeson production ($p\gamma$),  
Bethe--Heitler pair production ($p\gamma\rightarrow pe^{+}e^{-}$),  
and inelastic proton--proton collisions ($pp$).  
The $p\gamma$ cooling rate is~\cite{PhysRevLett.21.1016,PhysRevLett.78.2292,PhysRevD.76.123001}
\begin{equation}
t_{p,p\gamma}^{-1} = 
\frac{c}{2\gamma_p}
\!\int_{\bar{\epsilon}_{\mathrm{th}}}^{\infty}
\!\! d\bar{\epsilon}\,
\sigma_{p\gamma}(\bar{\epsilon})\kappa_{p\gamma}(\bar{\epsilon})\bar{\epsilon}
\!\int_{\bar{\epsilon}/(2\gamma_p)}^{\infty}
\!\! d\epsilon_\gamma\,\epsilon_\gamma^{-2}\frac{dn_\gamma}{d\epsilon_\gamma},
\label{eq:t_pgamma}
\end{equation}
where $\gamma_p=\epsilon_p/(m_p c^{2})$,  
$\bar{\epsilon}$ is the photon energy in the proton frame,  
$\bar{\epsilon}_{\mathrm{th}}\approx 145~\mathrm{MeV}$,  
and $dn_\gamma/d\epsilon_\gamma$ is the photon spectrum.
Cross sections ($\sigma_{p\gamma}$) and inelasticities ($\kappa_{p\gamma}$) are taken from Refs.~\cite{Patrignani_2016,article}. 

For the Bethe--Heitler process, the threshold energy is replaced by
$\bar{\epsilon}_{\mathrm{th,BH}}\simeq 1\mathrm{MeV}$. 
The rate $t_{\mathrm{BH}}^{-1}$ is then evaluated with 
$\sigma_{p\gamma}\to\sigma_{\mathrm{BH}}$ and 
$\kappa_{p\gamma}\to\kappa_{\mathrm{BH}}$~\cite{1992ApJ...400..181C}.

For $pp$ collisions, we use $t_{p,pp} = {1}/{(c\,\sigma_{pp}\,n_p\,\kappa_{pp})}$,
where $\sigma_{pp} \simeq 5 \times 10^{-26}\,\mathrm{cm^{2}}$ is taken from Ref.~\cite{2006PhRvD..74c4018K} and $\kappa_{pp}\simeq 0.5$. And the adiabatic cooling timescale is
$t_{p,\mathrm{adi}} = {r}/{(\beta c\,\Gamma)}$,
where $\beta$ and $\Gamma$ are the velocity (in units of $c$) and Lorentz factor of the relevant region.

Combining all channels, the total cooling rate is
\begin{equation}
t_{p,\mathrm{cool}}^{-1}
= t_{p,pp}^{-1} + t_{p,p\gamma}^{-1} + t_{p,\mathrm{BH}}^{-1}
+ t_{p,\mathrm{syn}}^{-1} + t_{p,\mathrm{IC}}^{-1}
+ t_{p,\mathrm{adi}}^{-1}.
\end{equation}
Since only $pp$ and $p\gamma$ interactions produce neutrinos,  
the proton suppression factor is
\begin{equation}
\zeta_{p,\mathrm{sup}}(\epsilon_{\nu_i})
= \frac{t_{p,pp}^{-1} + t_{p,p\gamma}^{-1}}
       {t_{p,\mathrm{cool}}^{-1}}.
\label{eq:zeta_p}
\end{equation}

Both $pp$ and $p\gamma$ interactions produce charged pions and kaons, which decay into muons and neutrinos.  
Muons further decay into electron and muon neutrinos.  
These intermediate particles may cool before they decay.  
For pions and kaons, the hadronic cooling timescale is
$t_{\{\pi,K\},\mathrm{had}}
=1/{(c\,\sigma_{\mathrm{h}}\,n_p\,\kappa_{\mathrm{h}})}$,
where $\sigma_{\mathrm{h}}\simeq 5\times10^{-26}\,\mathrm{cm^{2}}$  
and $\kappa_{\mathrm{h}}\simeq 0.8$~\cite{Olive_2014}.  
Hadronic cooling of muons is negligible.  
Synchrotron and IC cooling follow the same forms as Eqs.~\eqref{eq:t_syn}--\eqref{eq:t_IC} with the proton mass replaced by the mass of the relevant particle.

The decay timescale of particle $i=\pi,K,\mu_{\pi},\mu_{K}$ is 
$t_{i,\mathrm{dec}} = \gamma_i \tau_i$,
with $\gamma_i=\epsilon_i'/(m_i c^{2})$ and $\tau_i$ the rest-frame lifetime.  
The total cooling rate is
\begin{equation}
t_{i,\mathrm{cool}}^{-1}
= t_{i,\mathrm{dec}}^{-1} + t_{i,\mathrm{had}}^{-1}
+ t_{i,\mathrm{syn}}^{-1} + t_{i,\mathrm{IC}}^{-1}
+ t_{i,\mathrm{adi}}^{-1},
\end{equation}
with $t_{i,\mathrm{had}}=0$ for muons.  
The corresponding suppression factor is
\begin{equation}
\zeta_{i,\mathrm{sup}}(\epsilon_{\nu_i})
= \frac{t_{i,\mathrm{dec}}^{-1}}{t_{i,\mathrm{cool}}^{-1}}.
\end{equation}

We assume a proton spectrum $dN_p/d\epsilon_p\propto\epsilon_p^{-2}$.  
The minimum proton energy is $\epsilon_{p,\min}\approx \Gamma m_{p}c^{2}$,  
and we set $\epsilon_{p,\max}=10$~PeV~\cite{2021ApJ...911L..19Z}, as variations in $\epsilon_{p,\max}$ have a negligible impact on our results.  
The neutrino spectrum from a single channel $i$ is
\begin{equation}
\epsilon_{\nu_i}^{2}F_{\nu_i}
=
\frac{
N_i \, E_{\mathrm{p}}\,
\zeta_{p,\mathrm{sup}}(\epsilon_{\nu_i})\,
\zeta_{i,\mathrm{sup}}(\epsilon_{\nu_i})
}
{4\pi D_L^{2}
\ln\left(\epsilon_{p,\max}/\epsilon_{p,\min}\right)},
\end{equation}
where $E_{\mathrm{p}}$ is the total proton energy,
$N_{\pi}=N_{\mu_{\pi}}=0.12$,
$N_{K}=0.009$, and
$N_{\mu_{K}}=0.003$~\cite{Denton_2018,2021ApJ...911L..19Z}.
The neutrino energy is $\epsilon_{\nu_i}=a_i \Gamma \epsilon_p$, with
$a_{\pi}=a_{\mu_{\pi}}=0.05$,
$a_{K}=0.10$, and
$a_{\mu_{K}}=0.033$~\cite{Denton_2018,2021ApJ...911L..19Z}.

Integrating from the progenitor stellar radius $r_{\mathrm{ps}} = 10^{11}\,\mathrm{cm}$ to the jet choking radius $r_{\mathrm{ch}}$ gives the total fluence
\begin{equation}
\begin{aligned}
\epsilon_{\nu_i}^{2}F_{\nu_i,\mathrm{tot}}
&=
\frac{N_i}{4\pi D_L^{2}}
\int_{r_{\mathrm{ps}}}^{r_{\mathrm{ch}}} dr \,
\frac{4\pi r^{2} e(r)\,\xi_p}
{\ln\!\left[\epsilon_{p,\max}/\epsilon_{p,\min}(r)\right]}
\\
&\qquad\times
\zeta_{p,\mathrm{sup}}(\epsilon_{\nu_i},r)\,
\zeta_{i,\mathrm{sup}}(\epsilon_{\nu_i},r) ,
\end{aligned}
\end{equation}
where $\xi_p$ is the proton energy fraction.
Summing over all channels yields
\begin{equation}
\epsilon_{\nu}^{2}F_{\nu,\mathrm{tot}}
=
\epsilon_{\nu}^{2}F_{\nu_{\pi}}
+ \epsilon_{\nu}^{2}F_{\nu_{K}}
+ \epsilon_{\nu}^{2}F_{\nu_{\mu_{\pi}}}
+ \epsilon_{\nu}^{2}F_{\nu_{\mu_{K}}}.
\end{equation}

The energy dissipation discussed above occurs at the jet head; in addition, energy dissipation can also take place within the jet itself. 
For a non-magnetized jet, internal shocks dissipate part of the jet energy and also produce neutrinos.  
We assume a dissipation fraction $\iota = 0.1$~\cite{2018pgrb.book.....Z} at the radius  
$r_{\mathrm{is}}=\Gamma_j^{2}c\Delta t\approx10^{12}\,\mathrm{cm}$~\cite{1994ApJ...430L..93R,1998MNRAS.296..275D}.  
The internal-shock duration is much shorter than the jet-propagation time,  
so we treat this dissipation as instantaneous.  
After this episode, the intrinsic luminosity becomes $(1-\iota)L_0$.
The target photons for internal shocks originate from those advected from Region~III,  
boosted by a factor $\gamma_{34}$ in the shock frame.  
We include the photon escape fraction  
$f_{\mathrm{esc}}\sim(n_{\gamma,3}\sigma_{\mathrm{T}} r_{\text{rs}}/\gamma_3)^{-1}$~\cite{2016PhRvD..93h3003S}.  
The radiation constraint for proton acceleration then becomes~\cite{2013PhRvL.111l1102M}
\begin{equation}
L_{0,52}\, r_{\mathrm{IS},10}^{-1}\, \Gamma_2^{-3}
\lesssim 5.7\times10^{-3}
\min\!\left[\Gamma_{\mathrm{rel},0.5}^{2},\;
0.32\,C_1^{-1}\Gamma_{\mathrm{rel},0.5}^{3}\right],
\end{equation}
where $\Gamma_{\mathrm{rel}}$ is the relative Lorentz factor of the two shells,  
$\Gamma_{\mathrm{rel}}\simeq(\Gamma_{\mathrm{r}}/\Gamma_{\mathrm{s}}+\Gamma_{\mathrm{s}}/\Gamma_{\mathrm{r}})/2\sim3$~\cite{2018ApJ...856..119H},  
and $C=1+2\ln(\Gamma_{\mathrm{rel}}^{2})$.

For magnetized jets, internal shocks are not expected to efficiently dissipate energy for proton acceleration. 
Instead, the prompt emission is more naturally described by the ICMART model~\cite{2011ApJ...726...90Z}, while the associated neutrino production is expected to be negligible because the dissipation occurs far from the central engine, where the particle number density is low~\cite{PhysRevLett.110.121101,2023ApJ...944..115A,2024ApJ...976..174O,2025ApJ...987...79L,2025arXiv251022914O}. 
We therefore neglect the neutrino contribution from the ICMART process in this work.

\subsection{Density Profile of the Extended Material}
\label{sec:density_profile}

The density distribution of the extended material surrounding the progenitor plays a key role in determining the final fate of the jet and the efficiency of neutrino production.
However, the density structure of the outer layers of Wolf--Rayet stars remains poorly constrained.
We use GRB~060218 as a benchmark and adopt the density normalization inferred in Sec.~\ref{sec:jet_dynamics}, namely $\rho(r)\simeq6.3\times10^{13}\,r^{-2}\,\mathrm{g\,cm^{-3}}$ (corresponding to an exponent of $13.8$), as a characteristic reference value.

Instead of assuming a fixed normalization, we allow the density-normalization index to vary.
We model it as a Gaussian distribution centered at $13.8$.
The corresponding standard deviation is constrained by the observed event-rate ratio between low-luminosity and high-luminosity GRBs.

We associate the final fate of a propagating jet with the observed GRB phenomenology.
The jet evolution is characterized by three radii:
(i) the complete choking radius $r_{\mathrm{ch}}$, where the reverse shock has fully crossed the jet;
(ii) the prompt-emission radius $r_{\mathrm{pe}}$, which corresponds to the main dissipation site of prompt $\gamma$-ray emission; and
(iii) the shock-breakout radius $r_{\mathrm{sb}}$, where the forward shock emerges from the photosphere of the extended material.

Depending on the ordering of these radii, three distinct outcomes arise:

\begin{itemize}
    \item If $r_{\mathrm{ch}} < r_{\mathrm{sb}}$, the jet is completely choked inside the extended material and fails to emerge.
    We classify this case as an event associated with a fully choked jet, in which the jet fails to break out of the surrounding material.
    Such events can produce high-energy neutrinos but do not show a prompt X-ray or gamma-ray counterpart.

    \item If $r_{\mathrm{sb}} < r_{\mathrm{pe}} < r_{\mathrm{ch}}$, the jet successfully breaks out and produces a classical prompt-emission phase.
    These events are identified as HLGRBs.

    \item All remaining cases correspond to shock breakout without a successful prompt-emission phase.
    We classify these events as LLGRBs.
\end{itemize}

Observationally, the event-rate ratio between LLGRBs and HLGRBs is approximately $200{:}1$~\cite{2015ApJ...812...33S}.
To reproduce this ratio, we model a population of relativistic jets by sampling their intrinsic properties from observationally motivated distributions.

We assume that the luminosity function of HLGRBs traces the intrinsic distribution of jet luminosities.
We adopt the broken power-law luminosity function from Ref.~\cite{2015ApJ...812...33S}:
\begin{equation}
\begin{aligned}
\Phi(L) \propto 
\Bigg\{
&\Bigg[
\Bigg(
\left(\frac{L_{b,1}}{L_{b,2}}\right)^{\alpha_2}
+
\left(\frac{L_{b,1}}{L_{b,2}}\right)^{\alpha_3}
\Bigg)^{-1}
\left(\frac{L}{L_{b,1}}\right)^{-\alpha_1}
\Bigg]^{-1}
\\[4pt]
&
+
\Bigg[
\left(\frac{L}{L_{b,2}}\right)^{\alpha_2}
+
\left(\frac{L}{L_{b,2}}\right)^{\alpha_3}
\Bigg]^{-1}
\Bigg\},
\end{aligned}
\label{eq:LF}
\end{equation}
with parameters
\begin{equation}
\alpha_1 = 2.2^{+0.4}_{-0.2},\;
\alpha_2 = 1.0^{+0.1}_{-0.1},\;
\alpha_3 = 2.0^{+0.3}_{-0.3},
\end{equation}
and break luminosities
\begin{equation}
L_{b,1} = 5.0^{+3.0}_{-1.3}\times10^{50}\,\mathrm{erg\,s^{-1}},\quad
L_{b,2} = 7.1^{+4.3}_{-3.0}\times10^{52}\,\mathrm{erg\,s^{-1}}.
\end{equation}

The engine duration is sampled from the $T_{90}$ distribution of the \emph{Fermi}/GBM GRB sample~\cite{2013ApJ...763...15Q}.
The distribution of $\log T_{90}$ is well described by a Gaussian with a mean of $1.32$ and a standard deviation of $0.49$.

Recent multimessenger observations suggest that GRB jets are highly magnetized~\cite{2023ApJ...944..115A,2024ApJ...972....1L,2025arXiv251022914O}.
Time-resolved spectral analyses of \emph{Fermi}/GBM GRBs show that most prompt spectra are well described by power-law-like models.
This behavior is consistent with strongly magnetized outflows with $\sigma \gtrsim 10$~\cite{2024ApJ...972....1L}.
Motivated by these results, we focus on the magnetized scenario and adopt $\sigma=10$ in our simulations.

By fitting the simulated jet population to the observed LLGRB-to-HLGRB event-rate ratio, we find that a Gaussian dispersion of $0.56$ in the density-normalization index provides good agreement with the data in the magnetized case.

For comparison, we also perform the same calculation for the non-magnetized case. 
Adopting a density-normalization index centered at 13.8, we find that even with a standard deviation of 0, the non-magnetized population cannot reproduce the observed LLGRB-to-HLGRB event-rate ratio of $\sim200{:}1$. 
Reproducing such a ratio in the non-magnetized scenario instead requires a substantially higher density normalization; for the same standard deviation, the required central value must be increased by approximately an order of magnitude.

Taken together with the multimessenger evidence favoring magnetized GRB jets, we therefore adopt a Gaussian distribution for the density normalization of the extended material surrounding LLGRB progenitors, centered at 13.8 with a standard deviation of 0.56.

\subsection{Redshift Distribution}
\label{sec:redshift_distribution}

We next describe the redshift distribution of LLGRBs adopted in this work.

We model the redshift dependence of the LLGRB formation rate by assuming that it traces the cosmic star-formation rate~\cite{2008ApJ...683L...5Y},
\begin{equation}
f(z) \propto 
\left[
(1+z)^{p_1 \kappa}
+
\left(\frac{1+z}{5000}\right)^{p_2 \kappa}
+
\left(\frac{1+z}{9}\right)^{p_3 \kappa}
\right]^{1/\kappa},
\label{eq:Rz_LLGRB}
\end{equation}
where $\kappa=-10$, $p_1=3.4$, $p_2=-0.3$, and $p_3=-3.5$.

The comoving volume element is given by~\cite{2019ApJ...881L..40S}
\begin{equation}
\begin{aligned}
\frac{dV(z)}{dz}
&=
4\pi \left(\frac{c}{H_0}\right)^3
\left[
\int_0^z
\frac{dz}{\sqrt{1-\Omega_m+\Omega_m(1+z)^3}}
\right]^2 \\
&\quad\times
\frac{1}{\sqrt{1-\Omega_m+\Omega_m(1+z)^3}} .
\end{aligned}
\label{eq:dVdz}
\end{equation}
where we adopt the Planck cosmological parameters,
$H_0 = 67.8\,\mathrm{km\,s^{-1}\,Mpc^{-1}}$,
$\Omega_m = 0.308$, and $\Omega_\Lambda = 0.692$~\cite{2016A&A...594A..13P}.

Combining the LLGRB formation rate with the comoving volume element, the redshift
distribution of LLGRBs can be written as~\cite{2019ApJ...881L..40S}
\begin{equation}
N_{\rm LLGRB}(z)
=
\frac{R_{\rm LLGRB,0}\, f(z)}{1+z}
\frac{dV(z)}{dz},
\label{eq:Nz_LLGRB}
\end{equation}
where $R_{\rm LLGRB,0} \simeq 164\,\mathrm{Gpc^{-3}\,yr^{-1}}$ is the local event rate~\cite{2015ApJ...812...33S}.

This redshift distribution plays a key role in our calculations of joint-detection rates and the resulting diffuse neutrino flux.

\section{Prospects for Joint Observations}
\label{sec:joint_observation}

In this section, we investigate the prospects for multimessenger observations of LLGRBs, considering both single events and stacked samples. 
We focus on current and future high-energy neutrino observatories in combination with X-ray transient detectors.

\subsection{Single-Event Joint Observations}
\label{sec:single_event}

We first illustrate the single-event detection prospects using a GRB~060218-like LLGRB. 
Adopting the jet parameters defined in Sec.~\ref{sec:jet_dynamics} for an intrinsic luminosity of $L_0 = 10^{49}\,\mathrm{erg\,s^{-1}}$, we compute the expected neutrino spectra for both non-magnetized and magnetized outflows. 
\begin{figure}[htbp]
    \centering
    \includegraphics[width=0.45\textwidth]{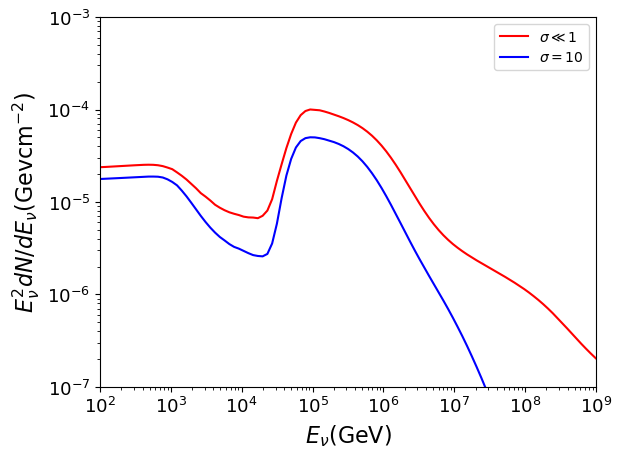}
    \caption{Neutrino energy spectrum for a GRB~060218–like event.  
    The red and blue curves correspond to different jet magnetization parameters, $\sigma \ll 1$ and $\sigma = 10$, respectively.
    GRB~060218 has an observed redshift of $z=0.033$~\cite{2006GCN..4792....1M}, which is used in our calculations.}
    \label{fig:060218_fluence}
\end{figure}
The resulting fluences are shown in Fig.~\ref{fig:060218_fluence}. 
Using the IceCube effective area~\cite{2021arXiv210109836I} evaluated at a source declination of $+16^{\circ}52'02.6''$~\cite{2006IAUC.8674....2S}, we find that the corresponding detection probabilities are extremely low, with $90\%$ detection probabilities of $\sim0.2\%$ and $\sim0.09\%$ for the non-magnetized and magnetized cases, respectively.
This example highlights the intrinsic difficulty of detecting neutrinos from individual LLGRBs with current instruments.

Motivated by this result, we next explore the joint-detection prospects in a more general and systematic manner. 
We examine the parameter space spanned by the isotropic-equivalent luminosity $L_{\mathrm{iso}}$ and the luminosity distance $D_L$ for a single LLGRB, and evaluate detectability with IceCube as well as with a next-generation neutrino telescope. 
For the neutrino side, we adopt an IceCube effective area~\cite{2021arXiv210109836I} averaged over the sine of the source declination, so that our results do not depend on the source declination.
We further account for uncertainties in the extended-material density and the engine duration by varying these parameters within one standard deviation of their respective distributions, as defined in Sec.~\ref{sec:theory}. 
\begin{figure}[htbp]
    \centering
    \textbf{Non-magnetized jet ($\sigma \ll 1$)} \\[3pt]
    \includegraphics[width=0.45\textwidth]{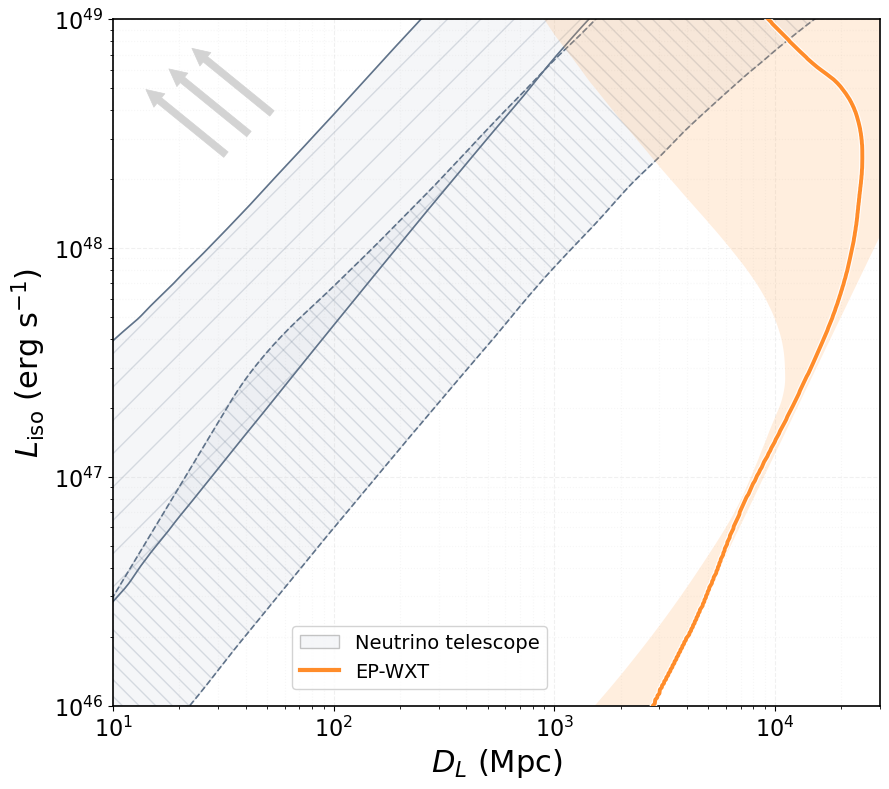} \\
    \vspace{10pt}
    \textbf{Magnetized jet ($\sigma=10$)} \\[3pt]
    \includegraphics[width=0.45\textwidth]{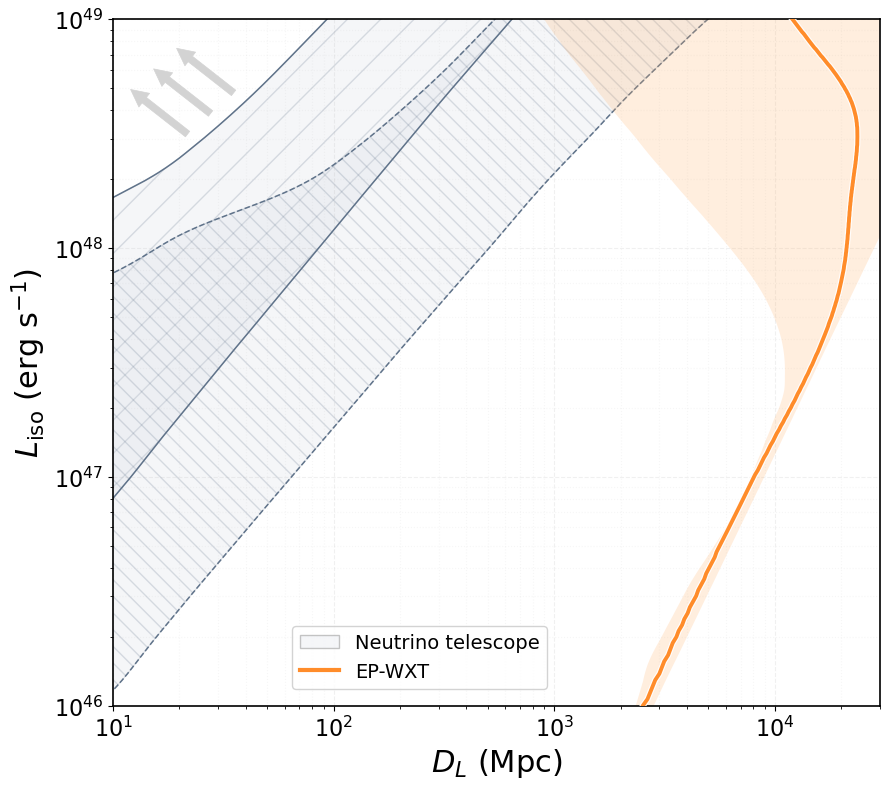}
    \caption{
    Parameter space in the $L_{\mathrm{iso}}$–$D_L$ plane for individual LLGRBs.
    Variations in the extended-material density and the engine duration at the one-standard-deviation level give rise to a finite width in both the electromagnetic and neutrino sensitivities shown here.
    The light-orange shaded region indicates the sensitivity range of the \emph{Einstein Probe} Wide-field X-ray Telescope, with the orange solid curve indicating the representative sensitivity (the median value).
    Hatched bands represent the corresponding $90\%$ neutrino detection limits.
    The band with left-slash hatching ($/$) corresponds to the IceCube sensitivity, while the band with right-slash hatching ($\backslash$) shows the sensitivity of a next-generation neutrino telescope with a sensitivity of $\sim30$ times that of IceCube.
    Sources located in the left-hand overlap region between the EP–WXT sensitivity range and a given neutrino band can be jointly detected.
    }
    \label{fig:lz_3}
\end{figure}
The resulting $90\%$ detection limits are shown as hatched bands in Fig.~\ref{fig:lz_3}.

For the X-ray counterpart, we adopt the \emph{Einstein Probe} Wide-field X-ray Telescope as a reference instrument, as it represents the most advanced wide-field X-ray transient detector currently planned. 
It operates in the $0.4$--$5~\mathrm{keV}$ energy range with a flux sensitivity of $10^{-11}\,\mathrm{erg\,cm^{-2}\,s^{-1}}$. We assume $L_X = L_{\text{iso}}$.
The orange solid line in Fig.~\ref{fig:lz_3} indicates the representative sensitivity.


For single events in the non-magnetized case ($\sigma \ll 1$), joint detection is possible only if the source lies in the left-hand overlap region between the X-ray sensitivity curve and the neutrino detection band(left-slash hatching). 
The lower-right boundary of the band corresponds to the optimistic case, while the upper-left boundary represents the conservative case.
In this regime, the neutrino sensitivity is the primary limiting factor. 
For example, for a source at $D_L = 10^2\,\mathrm{Mpc}$, a luminosity of $\sim3.8\times10^{48}\,\mathrm{erg\,s^{-1}}$ is required in the conservative case, while $\sim4.6\times10^{47}\,\mathrm{erg\,s^{-1}}$ suffices in the optimistic case. 
Conversely, for a source with $L_{\mathrm{iso}} = 10^{48}\,\mathrm{erg\,s^{-1}}$, the maximum detectable distance is only $\sim2.7\times10^{1}\,\mathrm{Mpc}$ in the conservative case or $\sim1.9\times10^{2}\,\mathrm{Mpc}$ in the optimistic case. 
From these results, we see that joint X-ray and neutrino detection is generally limited to nearby LLGRBs or those with high luminosities.
Using the luminosity function and redshift distribution introduced in Sec.~\ref{sec:theory}, we sample the source population and combine it with the event rate to compute how many events fall within the region of the $L_{\mathrm{iso}}$–$D_L$ plane accessible to joint observations. 
This leads to an waiting time of order hundreds of years for a single joint detection, even under optimistic assumptions.

With the coming of next-generation neutrino telescopes, such as IceCube-Gen2~\cite{2021JPhG...48f0501A}, KM3NeT~\cite{2016JPhG...43h4001A}, TRIDENT~\cite{2023NatAs...7.1497Y}, and the Hunt experiment~\cite{2024icrc.confE1080H}, neutrino detection capabilities will be significantly improved. 
The effective areas of these instruments are expected to be enhanced by factors ranging from $\sim8$ to $\sim30$ relative to IceCube~\cite{2025ApJ...987...79L}.
Given these developments, we consider an optimistic scenario of a next-generation neutrino telescope with an effective area enhanced by a factor of $\sim30$ relative to IceCube.
In this case, the improved neutrino sensitivity allows joint detection to be explored over a wider region of the $L_{\mathrm{iso}}$--$D_L$ parameter space. 
For a source with a typical isotropic luminosity of $L_{\mathrm{iso}} = 10^{47}\,\mathrm{erg\,s^{-1}}$, the maximum detectable distance is $\sim2.1\times10^{1}\,\mathrm{Mpc}$ in the conservative case or $\sim1.6\times10^{2}\,\mathrm{Mpc}$ in the optimistic case. 
Alternatively, if the source is placed at a luminosity distance of $D_L \sim 10^{3}\,\mathrm{Mpc}$, joint detection requires $L_{\mathrm{iso}}\gtrsim6.6\times10^{48}\,\mathrm{erg\,s^{-1}}$ in the conservative case and $L_{\mathrm{iso}}\gtrsim8.2\times10^{47}\,\mathrm{erg\,s^{-1}}$ in the optimistic case.
These results indicate that next-generation neutrino telescopes can substantially enhance the prospects for joint observations of individual LLGRBs, leading to an expected detection rate of approximately one event per year in the optimistic case.

We now turn to the magnetized scenario ($\sigma=10$), which is more strongly supported by recent multimessenger observations~\cite{2024ApJ...972....1L, 2025ApJ...987...79L,2025arXiv251022914O}. 
In this case, proton acceleration is less efficient and radiative cooling is stronger, leading to a lower neutrino yield and a much narrower jointly detectable region.

For individual magnetized LLGRBs, IceCube can detect only extremely nearby and luminous events. 
For example, at $D_L = 10^{2}\,\mathrm{Mpc}$, the required luminosity is $\sim1.2\times10^{48}\,\mathrm{erg\,s^{-1}}$ in the optimistic case, while in the conservative case it must exceed $\sim10^{49}\,\mathrm{erg\,s^{-1}}$, which is higher than what is commonly inferred for LLGRBs. 
Conversely, for a source with $L_{\mathrm{iso}} = 10^{48}\,\mathrm{erg\,s^{-1}}$, the maximum detectable distance is less than $10\,\mathrm{Mpc}$ in the conservative case, or $\sim8.6\times10^{1}\,\mathrm{Mpc}$ in the optimistic case. 
These constraints imply that joint observations of individual magnetized LLGRBs with current instruments are highly unlikely, even under optimistic case, implying an average waiting time of order one thousand years for a joint observation when combined with the event rate

For a next-generation neutrino telescope with an effective area enhanced by a factor of $\sim30$ relative to IceCube, the prospects improve substantially. 
For a source with a typical isotropic luminosity of $L_{\mathrm{iso}} = 10^{47}\,\mathrm{erg\,s^{-1}}$, joint detection in the optimistic case is possible out to a maximum distance of $\sim6.5\times10^{1}\,\mathrm{Mpc}$, while in the conservative case it is also restricted to distances less than $10\,\mathrm{Mpc}$.
Alternatively, if the source is placed at a luminosity distance of $D_L \sim 10^{3}\,\mathrm{Mpc}$, joint detection requires $L_{\mathrm{iso}}$ to exceed $10^{49}\,\mathrm{erg\,s^{-1}}$ in the conservative case and $L_{\mathrm{iso}}\gtrsim2.1\times10^{48}\,\mathrm{erg\,s^{-1}}$ in the optimistic case.
Overall, these results indicate that a next-generation neutrino telescope with an effective area enhanced by a factor of $\sim30$ expands the accessible region in the $L_{\mathrm{iso}}$–$D_L$ parameter space for joint observations of individual magnetized LLGRBs. 
Nevertheless, the allowed region remains limited, leading to an expected detection rate of $\sim0.5$ event per year in the optimistic case.

\subsection{Stacked Joint Observations}
\label{sec:stacking}

Building on the single-event analysis, we now investigate the prospects for joint X-ray and neutrino observations using stacked samples, assuming a next-generation neutrino telescope with a sensitivity enhanced by a factor of $\sim30$ relative to IceCube. 
Stacking multiple sources enhances the cumulative neutrino signal and allows contributions from intrinsically fainter LLGRBs to be included. 
\begin{figure}[htbp]
    \centering
    \includegraphics[width=0.45\textwidth]{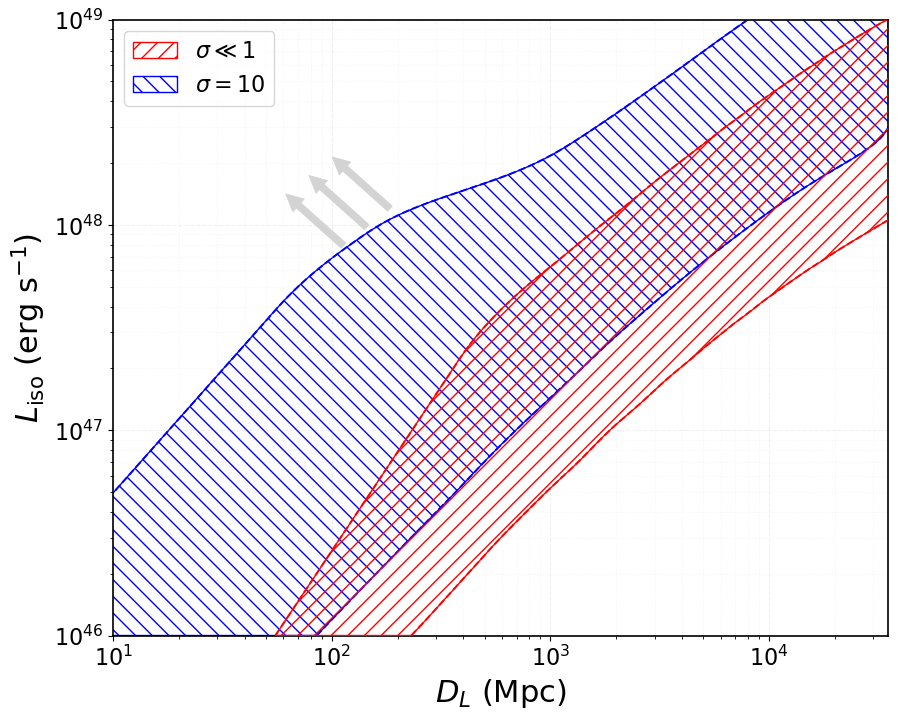} \\
    \vspace{10pt}
    \caption{
    Parameter space in the $L_{\mathrm{iso}}$–$D_L$ plane for stack LLGRBs.
    Variations in the extended-material density and the engine duration at the one-standard-deviation level give rise to a finite width in neutrino sensitivities shown here.
    Hatched bands represent the corresponding $90\%$ neutrino detection limits.
    The band with blue left-slash hatching ($/$) corresponds to the magnetized case, while the band with red right-slash hatching ($\backslash$) corresponds to the non-magnetized case.
    Sources located in the left-hand can be detected.
    }
    \label{fig:lz_3_1}
\end{figure}
As an illustrative assumption, we consider a stacked sample of $\sim100$ LLGRBs and adopt a simplified approach in which all sources are assumed to have identical intrinsic luminosities and distances. The corresponding neutrino constraints are shown in Fig.~\ref{fig:lz_3_1}.

We first summarize the non-magnetized case. 
For sources with typical luminosities of $L_{\mathrm{iso}} \sim 1\times10^{47}\,\mathrm{erg\,s^{-1}}$, stacking $\sim100$ LLGRBs allows one joint X-ray and neutrino detection to be achieved out to $D_L \lesssim 1.9\times10^{3}\,\mathrm{Mpc}$ in the optimistic case, while in the conservative case the accessible distance is reduced to $D_L \lesssim 2.2\times10^{2}\,\mathrm{Mpc}$. 
Alternatively, if the sources are placed at a luminosity distance of $D_L \sim 10^{3}\,\mathrm{Mpc}$, stacking enables joint detection for $L_{\mathrm{iso}}\gtrsim5.2\times10^{46}\,\mathrm{erg\,s^{-1}}$ in the optimistic case, while the conservative case requires $L_{\mathrm{iso}}\gtrsim6.3\times10^{47}\,\mathrm{erg\,s^{-1}}$.
In this scenario, stacking allows the contribution from much fainter LLGRBs to be included. 
The expected number of sources satisfying these conditions is of order $\sim400$ per year, allowing about four joint X-ray and neutrino detections to be achieved annually in the optimistic case.

We now also turn to the magnetized scenario.
For sources with a typical isotropic luminosity of $L_{\mathrm{iso}} \sim 1\times10^{47}\,\mathrm{erg\,s^{-1}}$, stacking a sample of $\sim100$ magnetized LLGRBs enables joint X-ray and neutrino detection out to $D_L \lesssim 7.0\times10^{2}\,\mathrm{Mpc}$ in the optimistic case, whereas in the conservative case the accessible distance is limited to $D_L \lesssim 1.7\times10^{1}\,\mathrm{Mpc}$. 
Alternatively, if the sources are placed at a luminosity distance of $D_L \sim 10^{3}\,\mathrm{Mpc}$, stacking enables joint detection for $L_{\mathrm{iso}}\gtrsim1.4\times10^{47}\,\mathrm{erg\,s^{-1}}$ in the optimistic case, while the conservative case requires $L_{\mathrm{iso}}\gtrsim2.2\times10^{48}\,\mathrm{erg\,s^{-1}}$.
In the optimistic case, the number of events satisfying these conditions is $\sim200$ per year, allowing about two joint X-ray and neutrino detections to be achieved annually in the optimistic case.

While we have also discussed stacked neutrino searches for the non-magnetized case, several observational considerations favor magnetized jets. 
Taken together, electromagnetic observations~\cite{2024ApJ...972....1L} and the persistent non-detection of neutrinos in temporal and spatial coincidence suggest that magnetized jets are common in GRBs~\cite{2023ApJ...944..115A,2025ApJ...987...79L,2025arXiv251022914O}. 
Consequently, stacked searches focusing on magnetized LLGRBs are particularly well motivated and may provide more meaningful constraints.

\section{Diffuse Neutrino Flux}
\label{sec:event_rate}

Motivated by the considerations discussed above, we focus on magnetized jets as the relevant source population contributing to neutrino emission from LLGRBs and choked-jet events. 
Based on the analysis in Sec.~\ref{sec:theory}, the intrinsic ratio of choked-jet events to observable LLGRBs in this scenario is approximately $1{:}2$, implying that a substantial fraction of jets are completely hidden from electromagnetic view. 
The ratio between the event rates of choked-jet events and LLGRBs is highly sensitive to the density distribution of the extended material surrounding the progenitor. 
At present, our knowledge of the extended material remains limited. 
A more accurate description of the density distribution in the future would allow tighter constraints on the fraction of choked-jet events.

We adopt a local LLGRB rate of $164\,\mathrm{Gpc^{-3}\,yr^{-1}}$~\cite{2015ApJ...812...33S}, which we associate with the shock-breakout event rate. 
Using the inferred ratio between choked-jet events and observable LLGRBs, we estimate the corresponding rate of choked-jet events in the magnetized scenario. 
Combining these rates with the single-source neutrino spectra and the distributions of the relevant model parameters derived in Sec.~\ref{sec:neutrino_emission}, we compute the total diffuse neutrino flux contributed by LLGRBs and choked-jet events.

\begin{figure}[t]
    \centering
    \includegraphics[width=0.85\linewidth]{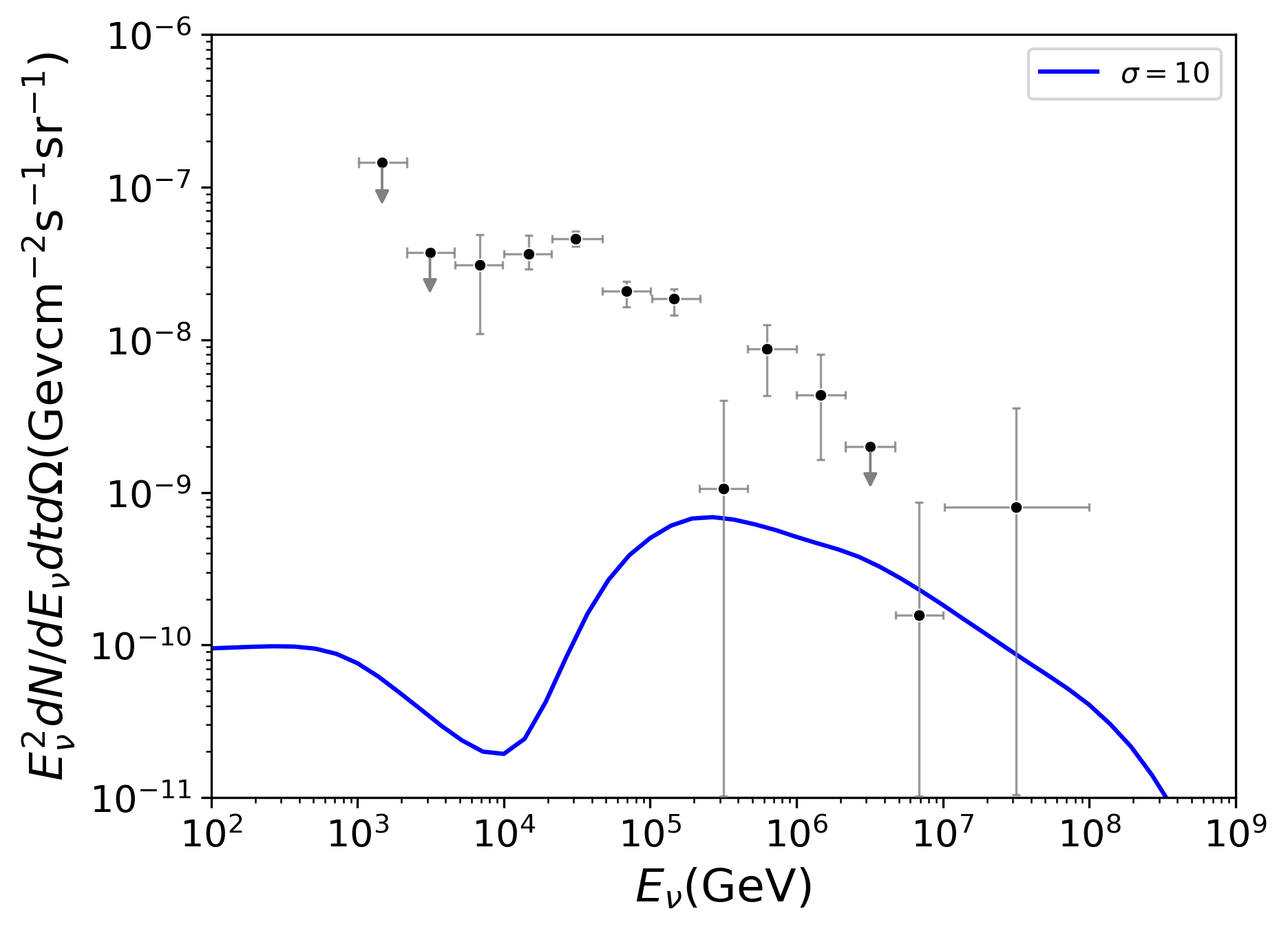}
    \caption{
        Predicted diffuse neutrino flux from shock-breakout events and choked-jet events in the magnetized scenario ($\sigma=10$).  
        Black points show the IceCube measurements of the diffuse neutrino flux~\cite{2025arXiv250722233A} for comparison.}
    \label{fig:diffuse}
\end{figure}

Figure~\ref{fig:diffuse} shows the resulting diffuse neutrino flux for $\sigma=10$. 
The predicted flux lies well below the IceCube measurements. 
We therefore conclude that LLGRBs and choked-jet events are unlikely to be the dominant contributors to the observed astrophysical diffuse neutrino flux.

\section{Conclusions}
\label{sec:conclusion}

We have presented a comprehensive study of high-energy neutrino production from LLGRBs.  
We developed a dynamical framework to describe the interaction between a relativistic jet and dense extended material surrounding a compact progenitor, and used it to compute the efficiency of neutrino production in different dynamical regimes.

By comparing three characteristic radii—the choking radius, the prompt-emission radius, and the shock-breakout radius—we classified jet outcomes into three categories: choked-jet events, HLGRBS, and LLGRBS.  

To connect with observations, we simulated a population of jets by sampling intrinsic luminosities, Lorentz factors, engine durations, and density normalizations from empirical or benchmark distributions.  
In the magnetized scenario with $\sigma=10$, our simulations reproduce the observed LLGRB to HLGRB event-rate ratio of $\sim 200:1$ with an appropriate choice of parameters.  
In this case we infer an intrinsic ratio of choked-jet events to LLGRBS of $\sim 1:2$, indicating that a substantial population of jets is completely hidden from electromagnetic observations.

Using the inferred event rates and the single-source neutrino spectra, we computed the diffuse neutrino flux contributed by shock-breakout events and choked jets.  
The predicted diffuse flux lies below the IceCube measurements.
We therefore conclude that these sources are unlikely to be the dominant origin of the astrophysical diffuse neutrino flux.

We assessed the prospects for joint X-ray and neutrino observations of LLGRBs. For individual events, joint X-ray and neutrino detection is generally limited to nearby LLGRBs or those with high luminosities.

For non-magnetized outflows ($\sigma \ll 1$), a next-generation neutrino telescope with an effective area $\sim30$ times larger than that of IceCube substantially improves the detectability of individual events.  
In the optimistic case, joint observation is enabled for sources with isotropic luminosities $L_{\mathrm{iso}}\sim1\times10^{47}\,\mathrm{erg\,s^{-1}}$ out to luminosity distances of $D_L \sim 1.6 \times 10^{2}\,\mathrm{Mpc}$. 
Alternatively, if the sources are placed at a luminosity distance of $D_L \sim 10^{3}\,\mathrm{Mpc}$, joint observation requires $L_{\mathrm{iso}}\gtrsim8.2\times10^{47}\,\mathrm{erg\,s^{-1}}$.
This corresponds to observing one joint event per year.

For magnetized outflows ($\sigma=10$), individual observation remains challenging even with enhanced sensitivity.  
However, stacking provides a viable pathway: for a stacked sample of $\sim100$ LLGRBs, one joint X-ray--neutrino observation can be achieved in the optimistic case for luminosities of $L_{\mathrm{iso}}\sim10^{47}\,\mathrm{erg\,s^{-1}}$ out to $D_L\lesssim 7.0 \times 10^{2}\,\mathrm{Mpc}$.  
Alternatively, if the sources are placed at a luminosity distance of $D_L\lesssim10^{3}\,\mathrm{Mpc}$, stacking enables joint detection for $L_{\mathrm{iso}}\gtrsim1.4\times10^{47}\,\mathrm{erg\,s^{-1}}$. 
In this case, the expected number of events satisfying these conditions is $\sim200$ per year, implying that joint X-ray and neutrino observations could be realized at a rate of order two events per year through stacking.

Our analysis shows that the primary limitation for multimessenger detections lies in the sensitivity of neutrino telescopes.
Next-generation high-energy neutrino telescopes will therefore be essential for uncovering and characterizing hidden, strongly magnetized jet populations and for advancing our understanding of jet physics and multimessenger transients.

\begin{acknowledgments}
We thank  Jinping Zhu, An Li, Jinda Li and Songyu Shen for helpful discussions and comments.  
This work was supported by the National Natural Science Foundation of China (Projects 12373040,12021003)
and the Fundamental Research Funds for the Central
Universities.
\end{acknowledgments}

\nocite{*}

\bibliography{apssamp}

\end{document}